\documentclass[a4paper, amsfonts, amssymb, amsmath, reprint, showkeys, nofootinbib, twoside, floatfix]{revtex4-1}

\usepackage[english]{babel}

\usepackage[utf8]{inputenc}
\usepackage{amsthm}
\usepackage{mathtools}
\usepackage{physics}

\usepackage{xcolor}
\usepackage{graphicx}
\usepackage[left=23mm,right=13mm,top=35mm,columnsep=15pt]{geometry} 
\usepackage{adjustbox}
\usepackage{placeins}
\usepackage[T1]{fontenc}
\usepackage{lipsum}
\usepackage{csquotes}
\usepackage{subfigure}
\usepackage{comment}

\newcommand{\rec}[1]{\frac{1}{#1}}

\newcommand{\vtau}{\vb*{\tau}}
\newcommand{\gratau}{\grad_{\vtau}}

\newcommand{\conj}[1]{#1^{*}}

\newcommand{\cev}[1]{\ev{#1}_E}

\newcommand{\eps}{\varepsilon}

\usepackage[pdftex, pdftitle={Article}, pdfauthor={Author}]{hyperref}

\begin{document}
\title{Conditional wave theory of environmental interactions with a quantum particle}

\author{Rory van Geleuken}
\author{Andrew V. Martin}
    \email[Correspondence email address: ]{andrew.martin@rmit.edu.au}
    \affiliation{School of Science, RMIT University, Melbourne, Victoria 3000, Australia.}

\date{July 2020}

\begin{abstract}
We present an alternative formulation of quantum decoherence theory using conditional wave theory (CWT), which was originally developed in molecular physics (also known as exact factorisation methods). We formulate a CWT of a classic model of collisional decoherence of a free particle with environmental particles treated in a long-wavelength limit. In general, the CWT equation of motion for the particle is non-linear, where the non-linearity enters via the CWT gauge fields. For Gaussian wave packets the analytic solutions of the CWT equations are in exact agreement with those from the density matrix formalism. We show that CWT gauge terms that determine the dynamics of the particle's marginal wave function are related to a Taylor series expansion of the particle's reduced density matrix. Approximate solutions to these equations lead to a linear-time approximation that reproduces the ensemble width in the limits of both short and long times, in addition to reproducing the long-term behaviour of the coherence length. With this approximation, the non-linear equation of motion for the particle's marginal wave function can be written in the form of the logarithmic Sch{\"o}dinger equation. The CWT formalism may lead to computationally efficient calculations of quantum decoherence, since it involves working with wave-function level terms instead of evolving a density matrix via a master equation.
\end{abstract}


\maketitle

\section{Introduction}

The interactions of quantum systems with their environments is of critical importance for the development of quantum technologies and for our understanding of the quantum world \cite{Joos2003,Schlosshauer2007}. However, environments contain large numbers of particles for which it is not possible to solve the quantum equations of motion exactly. Hence, models of environmental interactions are necessarily approximate. In many contexts, approximate models of environmental interactions treat decoherence as the primary effect of an environment on a quantum system \cite{Joos2003,Schlosshauer2007}. Decoherence involves the loss of coherence of a quantum system as it becomes entangled with the quantum states of the environment. Decoherence is of key importance in the development of quantum technologies such as quantum computing \cite{Chuang1995} and quantum communications \cite{Manz2007}, because it sets limits on performance and feasibility.

The common treatment of environmental interactions involves modelling the quantum system of interest as an open quantum system \cite{Joos2003,Schlosshauer2007}. The state of an open quantum system is modelled by a reduced density matrix $\rho(\vb{r}, \vb{r}', t)$ and its equation of motion is known as a master equation. Master equations have been modelled for a large class of quantum systems including free particles \cite{Joos1985}, spin systems \cite{Leggett1987} and quantum computers \cite{Chuang1995}. One challenge of the master equations approach is that the density matrix is of the size of the quantum system squared. Due to this scaling, additional numerical approximations, such as perturbation theory or Monte-Carlo approaches \cite{Zhang2007}, are usually required to solve master equations for the time-evolution of the quantum system.

Here we present an alternative approach to modelling quantum decoherence based on the technique of exact factorisation \cite{Hunter1975,Abedi2012}, which has been developed in the context of molecular physics. In that context it has been further developed to model various processes including light-matter interactions \cite{Hoffmann2018}, non-adiabatic dynamics \cite{Min2015} and electronic decoherence \cite{Curchod2016}. In electron microscopy a similar framework has been used to develop a quantum description of phonon scattering \cite{Forbes2010}. Our goal is to develop a description of collisional decoherence of a free-particle at a wave function level and explore an alternative to the use of density matrices.  

The factorization of the quantum state is known to many as the starting point of the Born-Oppenheimer approximation for the quantum description of a molecule:
\begin{equation} \label{factors0}
		\Psi(\vtau, \vb{r}, t) = a(\vtau, t)\phi(\vtau,\vb{r},t) \;.
\end{equation}
In molecular physics, $\vtau$ are the nuclear degrees of freedom in the molecule and $\vb{r}$ are the electronic degrees of freedom. In the context of quantum decoherence, we will use $\vtau$ to denote the spatial coordinates of a system of interest, and $\vb{r}$ to denote the environmental degrees of freedom. The reinterpretation of coordinates is possible because Eq. \eqref{factors0} is a general mathematical statement without restriction to any particular physical system. The factorization is potentially useful when the complexity of a quantum system is assigned to the conditional state, $\phi(\vtau,\vb{r},t)$, and appropriate approximations found to make calculations tractable. In the case of molecular physics, the complexity arises from the coupling between the electronic and nuclear degrees of freedom, where Eq. \eqref{factors0} facilitates the Born-Oppenheimer approximation. In the context of decoherence, $\phi(\vtau,\vb{r},t)$ describes the complexity of the environment. The concept of a conditional environmental state was already used in early derivations of collisional decoherence \cite{Joos1985, Schlosshauer2007}, which suggests that the concepts of exact factorization may carry over quite naturally to quantum decoherence. 

It has been recognised that with appropriate normalization conditions that $|a(\vtau, t)|^2$ and $|\phi(\vtau,\vb{r},t)|^2$ can be interpreted as a marginal and conditional probability densities respectively \cite{Hunter1975}. The normalization condition does not uniquely specify the phase of $a(\vtau, t)$ and $\phi(\vtau,\vb{r},t)$ and a redefinition of the phase gives rise to gauge terms in the equations of motion for both factors \cite{Abedi2012}. Despite its use in Born-Oppenheimer theory, the factorization itself does not involve an approximation. The Born-Oppenheimer approximation is a modification to the equations motion to make the numerical evaluation of the equation of motion for $\phi$ tractable. Attempts to solve the full gauge theory without the BO approximation are known as the `exact factorization' techniques primarily because they avoid the BO approximation. Here though we prefer to emphasise the underlying probabilistic concepts and refer to theories deriving from Eq. \eqref{factors0} as conditional wave theories (CWT).

Here we identify the relationships between the density matrix formalism of decoherence and the CWT formalism. We show that the CWT gauge terms are related to the first-order terms in a Taylor series expansion of the density matrix. The higher order terms in the expansion are related to higher-order spatial derivatives of the environment's conditional state. This has important consequences for the time-evolution of the CWT gauge terms, which are part of infinite series of coupled equations. These coupled equations are equivalent to the coupled time-evolution of the expansion coefficients of the density matrix. Hence, without approximation, the CWT formalism treatment of decoherence involves the same number of equations as the density matrix formalism.

We then revisit a model system for decoherence of a free particle interacting with environmental scattering particles first introduced by Joos and Zeh \cite{Joos1985} (here referred to as the JZ model). We derive the analytic result for the time-evolution of a particle with a Gaussian distribution in CWT and show that it is agrees exactly with the JZ solution. 

Using CWT, we introduce approximations to the gauge terms with a linear time-dependence that correctly produce the short and long time limits of the JZ Gaussian solution. The linear approximation may facilitate extensions of CWT to non-Gaussian wave functions and lead to computationally efficient methods for simulating open quantum systems.

We further show that with the linear approximation the CWT marginal equation of motion takes the form of a logarithmic Schr{\"o}dinger equation with a non-linearity that itself increases linearly with time. The logarithmic Schr{\"o}dinger equation has been used approximately to describe non-linear optics \cite{Buljan2003} and nuclear physics \cite{Hefter1985}. It has bounded solutions if the sign of the non-linear term is negative \cite{Bialynicki1979}. We show the sign is positive in the CWT of quantum decoherence and leads to unbounded solutions of the logarithmic Schr{\"o}dinger equation, which ensures agreement with the predictions of existing decoherence theory. It is of fundamental interest that the logarithmic Schr{\"o}dinger equation appears in the CWT of quantum decoherence and may lead new understanding about the dynamics induced by decoherence processes.

\section{Theory: Equations of Motion}

Consider the total wavefunction, $\Psi$ of a particle with position $\vb*{\tau}$, and its environment with degrees of freedom described by a vector $\vb{r}$. It can be written as the product of two functions,
	\begin{equation} \label{factors}
		\Psi(\vtau, \vb{r}, t) = a(\vtau, t)\phi(\vtau,\vb{r},t) \;,
	\end{equation}
where $a$ is called the marginal wavefunction and $\phi$ is the conditional wavefunction.
To ensure that $|a|^2$ and $|\phi|^2$ can be interpreted in terms marginal and conditional probability respectively, the following normalization condition is required:
    \begin{equation} \label{phinorm}
        \int \dd{\vb{r}} \abs{\phi(\vtau,\vb{r}, t)}^2 = 1.
    \end{equation}
Note that the marginal wavefunction alone is sufficient to calculate the expectation value of the position operator, but this is not necessarily true of other observables. This is because the factorization is made in a particular basis, which here is real-space position. For example, if we compute the expectation value of the particle's momentum and enforce (\ref{phinorm}), we find,
    \begin{equation}
        \expval{\vb{p}} = \\ -i\hbar \int \dd{\vtau} \conj{a}(\vtau) \left( \gratau + \int \dd{\vb{r}} \conj{\phi}\gratau \phi\right) a(\vtau) \;,
    \end{equation}
which clearly depends on both the marginal and conditional wavefunctions. Furthermore, defining a vector field,
    \begin{equation}
        \vb{A}(\vtau,t) = -i\hbar \int \dd{\vb{r}} \conj{\phi}(\vtau,\vb{r},t)\gratau\phi(\vtau,\vb{r},t) \;,
    \end{equation}
we see that replacing the standard momentum operator with a modified version,
    \begin{equation}
        \vb{p} = -i\hbar \gratau \quad \rightarrow \quad \vb{p} = -i\hbar \gratau + \vb{A}(\vtau,t)
    \end{equation}
yields the standard expression for the momentum expectation value.

The equations of motion for the marginal and conditional wavefunctions were first derived in gauge invariant form by Ref. \cite{Abedi2012} and here we quote the key results. The Hamiltonian, $H$, for the entire systems can be written as
    \begin{equation}
        H = H_{\mathrm{int}} + H_{\mathrm{env}} - \frac{\hbar^2}{2m}\laplacian_{\vtau} \;,
    \end{equation}
where $H_{\mathrm{env}}$ acts purely on the environment and $H_{\mathrm{int}}$ describes the coupling between the particle and environment. The equations of motion for the marginal and conditional wave functions are given by
    \begin{align}
        i\hbar \pdv{a(\vtau,t)}{t} &= \rec{2m} D^{2}_{+} a(\vtau,t) + \eps(\tau,t)a(\vtau,t) \label{aeom} \\
        i\hbar \pdv{\phi(\vtau,\vb{r},t)}{t} &= H_\phi \phi(\vtau,\vb{r},t) - \eps(\tau,t)\phi(\vtau,\vb{r},t) \;, \label{phieom}
    \end{align}
where
    \begin{align}
        \vb{D}_\pm &= -i\hbar\gratau \pm \vb{A} \\
        \label{Hphi}
        H_\phi &= H_{\mathrm{int}} + H_{\mathrm{env}} + \rec{2m} D^{2}_{-} + \rec{m} \frac{\vb{D}_+a(\vtau,t)}{a(\vtau,t)}\vdot \vb{D_{-}}
    \end{align}
and
    \begin{equation}
        \eps(\vtau,t) = \int \dd{\vb{r}} \conj{\phi}[H_\phi - i\hbar \partial_t]\phi.
    \end{equation}
If the effective Hamiltonian, given by (\ref{Hphi}), were to be implemented in a numerical scheme, the final term would require some form of regularisation around zeroes in the marginal wavefunction. Indeed, zeroes of the total wavefunction are of significance in CWT, as the vanishing of either the conditional or marginal wavefunction implies the factorisation in (\ref{factors}) is ill-defined. However, in this paper, we will be considering an analytical Gaussian solution for the marginal wavefunction so this term is well-defined everywhere.

In this paper, we do not explicitly calculate the conditional wavefunction, instead focusing on the behaviour of the marginal wavefunction under general assumptions about the environment.

In using CWT to model decoherence, one of our goals is to write down relationships between the potentials, $\eps(\vtau,t)$ and $\vb{A}(\vtau,t)$, and the decoherence terms, such as the density matrix and decoherence rate parameters. In this way, we aim to use the marginal wave equation to model the effect of the environment on the dynamics of the particle's spatial probability distribution. Therefore, we can calculate the dynamical consequences of decoherence at the ``wavefunction level''. 

\subsection{Gauge Symmetry}
The factorisation given in (\ref{factors}) and constrained by (\ref{phinorm}) is not unique. We can multiply each factor by reciprocal phase factors like so,
    \begin{align}
        a(\vtau, t) &\rightarrow a(\vtau,t)e^{i\theta(\vtau,t)} \\
        \phi(\vtau, \vb{r}, t) &\rightarrow \phi(\vtau, \vb{r}, t)e^{-i\theta(\vtau, t)},
    \end{align}
where $\theta(\vtau,t)$ is an arbitrary real function, which leaves the total wavefunction $\Psi$ unchanged. As described in Ref. \cite{Abedi2012}, this represents a gauge freedom which transforms the quantities $\vb{A}$ and $\eps$ as
    \begin{equation}
        \vb{A}(\vb*{\tau},t) \rightarrow \vb{A}(\vb*{\tau},t) - \hbar \gratau \theta(\vtau,t)
    \end{equation}
and
    \begin{align}
        \eps(\vb*{\tau},t) \rightarrow \eps(\vb*{\tau},t) - \hbar \partial_t \theta(\vtau, t).
    \end{align}

\section{Relationship to the density matrix}
\label{sec:relationship_to_density_matrix}

In order to formulate quantum decoherence theory in terms of conditional wave theory, we construct the density matrix of the entire system from Eq. (\ref{factors}),
    \begin{equation}
        \rho(\vtau,\vtau';\vb{r},\vb{r}';t) = \conj{a}(\vtau', t)a(\vtau, t)\conj{\phi}(\vtau',\vb{r}',t)\phi(\vtau,\vb{r},t).
    \end{equation}
The reduced density matrix of the particle $\rho_S=\tr_E{\rho}$ is obtained by taking the trace over the environmental degrees of freedom,
    \begin{align}
    \begin{split} \label{ptrace}
        \rho_S(\vtau,\vtau') &= \conj{a}(\vtau')a(\vtau) \int \dd{\vb{r}} \conj{\phi}(\vtau',\vb{r})\phi(\vtau,\vb{r}) \\
        &= \rho_M(\vtau, \vtau') K(\vtau',\vtau) \;,
    \end{split}
    \end{align}
where we have defined the marginal density matrix, $\rho_M(\vtau,\vtau') = \conj{a}(\vtau')a(\vtau)$ and the function $K(\vtau, \vtau')$, which is given by,
    \begin{equation} \label{Kdef}
        K(\vtau, \vtau') = \int \dd{\vb{r}} \conj{\phi}(\vtau', \vb{r})\phi(\vtau, \vb{r}).
    \end{equation}
Equation (\ref{ptrace}) tells us that the effect of tracing over the environment is captured by multiplying the isolated marginal density matrix by  the integral $K(\tau,\tau')$, which is determined by the conditional wavefunction. 

This integral can be expanded in powers of $\vb{y} = \vtau-\vtau'$ using a Taylor series,
    \begin{align} \label{eq:K_expansion}
    \begin{split}
        K(\vtau, \vtau') &= \sum_{n=0}^{\infty} \frac{(\vtau-\vtau')^n}{n!} \int \dd{\vb{r}} \conj{\phi}(\vtau',\vb{r}) \grad_{\vtau'}^n \phi(\vtau',\vb{r}) \\
        & = \sum_{n=0}^{\infty} \eval{g_{(0,n)}\left(\tfrac{\vb{z}}{2}\right)}_{y=0}\frac{\vb{y}^n}{n!} \;,
    \end{split}
    \end{align}
where $\vb{z} = \vtau + \vtau'$.   
    
In general, $g_{(n,m)}(\vtau)$ is given by,
    \begin{equation} \label{gdef}
        g_{(n,m)}(\vtau) = \int \dd{\vb{r}} (\gratau^n\conj{\phi})\vdot(\gratau^m\phi).
    \end{equation}
For each $N=n+m$ there are only $N$ independent real scalar functions $g_{(n,m)}(\vtau)$ (see Appendix A for a proof). These independent terms can be chosen to be the functions $g_{(0,n)}(\vtau)$ that appear in Eq. \eqref{eq:K_expansion}. In this way, the particle's reduced density matrix can be constructed solely from the marginal wave function and the $g_{(0,n)}(\vtau)$ functions.

The gauge field itself can be found by differentiation of $K$,
    \begin{equation} \label{AfromK}
         \eval{\pdv{K}{y}}_{y=0} = \frac{i}{\hbar}\vb{A}(\vtau,t),
    \end{equation}
which is clearly proportional to $g_{(0,1)}(\vtau)$.

It is also possible to give a physical interpretation to the g-functions. Firstly, consider the kinetic energy term for the marginal wavefunction in a gauge where $A$ vanishes,
    \begin{equation}
        -\frac{\hbar^2}{2m}\int \dd{\vb{r}} \conj{\phi} \gratau (a\phi) = -\frac{\hbar^2}{2m} \left(\gratau a + g_{(0,2)}(\vtau)a\right).
    \end{equation}
The first term on the RHS is the usual kinetic energy of the marginal system, but the second term arises purely due to the coupling with the environment. We can thus think of $g_{(0,2)}(\tau)$ as a measure of the kinetic energy invested in entanglement with the environment.

More generally, looking at the definition of $K(\tau,\tau')$, equation (\ref{Kdef}), we see that it can be written as,

    \begin{align}
        K(\vtau,\vtau') &= \int \dd{\vb{r}} \conj{\phi}(r,\vtau') e^{\frac{i}{\hbar}(\vtau-\vtau')\cdot \hat{\vb{p}}}\phi(r,\vtau') \\
            &= \expval{\hat{T}(\vtau'-\vtau)}_E \;,
    \end{align}
where $\expval{...}_E$ denotes the conditional expectation value, and $\hat{T}(\vb{x})$ is the translation operator over a displacement $\vb{x}$. So $K(\vtau,\vtau')$ quantifies the coherence between different particle positions due to entanglement with the environment. It can be interpreted as the projection of the environmental state given some particle position $\vtau$ on the environmental state corresponding to a particle position $\vtau'$. Depending on the degree of entanglement with the environment, $K$ would then encode the degree of sensitivity of the environmental state on the particle state. For example, in the limiting case where conditional wavefunction is completely independent of the particle position, $K(\vtau,\vtau')$ is a constant function, and the particle is in a pure state (full coherence). If the conditional wavefunctions of two particle positions are completely orthogonal, then we might have,
\begin{equation}
    K(\vtau,\vtau') = \delta(\vtau-\vtau') \;,
\end{equation}
and the particle would be completely decohered in the position basis. In more realistic settings, such as the Gaussian case considered in Section \ref{sec:GaussianCase}, there is a finite, time-dependent coherence length.

\section{Derivation of the gauge field from density matrix formalism}
\label{sec:gauge_from_density_matrix}

The Joos-Zeh master equation (JZME) \cite{Joos1985} is a well-studied and general form that has been used to model the decoherence of a variety of systems\cite{Schlosshauer2007}. Restricting our attention to one dimension and defining the rotated coordinates $y=\tau-\tau'$ and $z=\tau+\tau'$, then the JZME takes the form
    \begin{equation}\label{JZME}
        i\hbar \dv{\rho_S}{t} = -\frac{2\hbar^2}{m} \pdv[2]{\rho_S}{y}{z} - i\Lambda y^2 \rho_S.
    \end{equation}
This can be used as a starting point for constructing conditional and marginal wavefunctions. Defining,
    \begin{align}
        a(\tau,t) &= e^{r(\tau,t)} \\
        K(\tau, \tau',t) &= e^{\gamma(\tau, \tau',t)},
    \end{align}
then the reduced density matrix can be written as,
    \begin{equation}
        \rho_S(\tau, \tau',t) = e^{r(\tau,t)+\conj{r}(\tau', t)+\gamma(\tau,\tau',t)}.
    \end{equation}
Expanding $r(\tau)=r((z+y)/2)$, $\conj{r}(\tau')=\conj{r}((z-y)/2)$, and $\gamma(\tau,\tau') = \gamma(y,z)$ in Taylor series around $y=0$,
    \begin{align}
        r(\tau) &= \sum_{n=0}^{\infty} r_n(z) \frac{y^n}{n!} \\
        \conj{r}(\tau') &= \sum_{n=0}^{\infty} (-1)^n\conj{r}_n(z)\frac{y^n}{n!} \label{rconjtaylor}\\ 
        \gamma(y,z) &= \sum_{n=0}^{\infty} \gamma_n(z) \frac{y^n}{n!},
    \end{align}
where $r_n(z)=\eval{\partial^{n}_y r(\tau)}_{y=0}$ and $\gamma_n(z)=\eval{\partial^{n}_y \gamma(y,z)}_{y=0}$. The $(-1)^n$ term in the summand of (\ref{rconjtaylor}) occurs because the argument on the left hand side is $\tau'=(z-y)/2$ and not $\tau=(z+y)/2$ as in the previous line.

By defining $Q(\tau,\tau') = r(\tau) + \conj{r}(\tau')$ the terms in the JZME corresponding to the kinetic energy and coupling of the marginal and conditional components can be seen clearly,
    \begin{align}
    \begin{split}
        i\hbar(\dot{Q} + \dot{\gamma}) = -\frac{2\hbar^2}{m} & \left[ Q_{,yz} + Q_{,y}Q_{,z} \right. \\
                                                                   & + Q_{,y}\gamma_{,z} + Q_{,z}\gamma_{,y} \\
                                                                   & + \left. \gamma_{,yz} + \gamma_{,y}\gamma_{,z} \right] \\
                                                                 - & i\Lambda y^2.
    \end{split}
    \end{align}
We then define $Q_n(z) = r_n(z) + (-1)^n \conj{r}_n(z)$ so that the above can be rewritten (for arbitrary $n$),
    \begin{align} \label{Qseries}
    \begin{split}
        \dot{Q}_n + \dot{\gamma}_n = -\frac{2i\hbar}{m} & \left( Q'_{n+1} + \gamma'_{n+1} \right) \\
                 - \frac{2i\hbar}{m} & \sum_{k=0}^{n} \binom{n}{k} \left( \gamma'_k Q_{n-k+1} + \gamma_{k+1} Q'_{n-k} \right. \\
                 & \qquad \qquad  \left. + Q'_{n-k} Q_{k+1} + \gamma'_{n-k} \gamma_{k+1}  \right) \\
                - \frac{2\Lambda}{\hbar} & \delta_{n2} \;,
    \end{split}
    \end{align}
where $\delta_{ij}$ is the Kronecker delta and primes denote differentiation with respect to $z$.

Equation \eqref{Qseries} has a clear interpretation that depends on the order of $n$. The $n=0$ equation describes the time evolution of the particle's spatial density, since $\gamma_0 = 1$ due to normalization. The $n=1$ equation describes the time-evolution of the phase of the marginal wave function and the gauge term $A(\tau)$, which is proportional to $\gamma_1$. The choice of the gauge condition is required to solve the $n=1$ equation. The terms $Q_n$ for $n \ge 2$ are spatial derivatives of $Q_0$ and $Q_1$ and, hence, are not independent. Hence, the equations for $n \ge 2$ provide the time evolution of the terms  $\gamma \ge 2$, which describe the entanglement between the particle and the environment. In general, we see that each order $n$ depends on the next order $n+1$ and so in general we would need to solve an infinite sequence of partial differential equations to solve for the marginal wavefunction and the gauge terms. 

However, in the JZ model, the source term for decoherence only appears in the $n=2$ equation and the coupling to terms of $n > 2$ occurs via the particle's kinetic energy term. This suggests there could be significant computational advantages to approximately truncating the series of equations at $n=2$, such that only three functions of the size of the wave-function need to be modelled instead of the full density matrix. In the next section we consider the analytic Gaussian solutions to the JZ model for which the truncation at $n=2$ is exact. This is justified, since it was shown by Joos and Zeh themselves that their master equation possessed analytic Gaussian solutions \cite{Joos1985}.

It is also possible to apply a similar analysis to the case of quantum Brownian motion, where the master equation is of the form,
    \begin{align}
    \begin{split}
        \dot{\rho_S} = & -\frac{2i\hbar}{m} \pdv[2]{\rho_S}{y}{z} - 2iM\hbar\omega^2y\rho_S  
        \\ \quad & + Cy\pdv{\rho_S}{y} + i F y\pdv{\rho_S}{z} 
        \\ & - \Lambda y^2\rho_S \;,
    \end{split}
    \end{align}
where $\omega$ is the temperature-dependent Lamb-shifted frequency of the central oscillator, $C$ parameterises dissipation due to the environmental interaction, $F$ is the so-called \textit{anomalous-diffusion coefficient}, and $\Lambda$ is the long-wavelength decoherence rate identical to the corresponding term in the JZME. These terms do not produce source terms higher than second order and so the corresponding equations of motion possess Gaussian solutions. 

However, if a general master equation were employed, such as the Gallis-Fleming form\cite{Gallis1990} - to which the JZME is a long-wavelength approximation - there would be source terms at all orders of $n$.

We also note that the $\varepsilon$ term appearing in the CWT equations of motion does not appear explicitly in Eq. \eqref{Qseries}. This is because equations of motion for $a$ and $a^*$ are added to generate Eq. \eqref{Qseries}, resulting in the a term proportional to $\varepsilon(\tau)-\varepsilon(\tau')$ which accounts for the system-environment entanglement and interaction terms of the master equation (that is, those terms which do not arise from kinetic energy of the system alone). Thus $\varepsilon$ would, in this case, be derived from the form of the master equation and not the other way around.

\section{Gaussian Solutions to the Equation of Motion} \label{sec:GaussianCase}

We now consider the analytic Gaussian solutions for the JMZE derived by Joos and Zeh. We show that such solutions also exist for the marginal and conditional wavefunctions and are in exact agreement with Joos' and Zeh's calculation using the reduced density matrix. We then consider a linear time approximation to these exact solutions that could lead to numerically efficient simulations of non-Gaussian cases.

\subsection{Agreement with Joos-Zeh Solution}

 If we work in a gauge where $A(\tau)$ vanishes, we know from (\ref{AfromK}) that,
    \begin{equation}
        \eval{\pdv{K}{y}}_{y=0} = 0.
    \end{equation}
We also observe that,
    \begin{align}
    \begin{split}
        \eval{\pdv{K}{z}}_{y=0} &= \eval{\left( \pdv{K}{\tau} + \pdv{K}{\tau'} \right)}_{\tau=\tau'} \\
        &= \frac{i}{\hbar} (A(\tau')-A(\tau')) = 0 \;,
    \end{split}
    \end{align}
which implies
    \begin{equation}
        \pdv{K}{z} \sim O(y) \;,
    \end{equation}
since if the partial derivative of $K$ with respect to $z$ was independent of $y$, evaluating it at $y=0$ would yield a function of $\eval{z}_{y=0}=2\tau$. Moreover, since we aim to reproduce the Gaussian solutions of Joos and Zeh, we can ignore terms in the exponent of $K$ of order greater than 2 in $y$ or $z$. Thus we are left with a $\ln K$ that is proportional to $yz$ or $y^2$. Thus we write
    \begin{equation}
        K(\tau, \tau') = e^{-\gamma_{JZ}(t)\frac{y^2}{2} + i\eta(t)yz} \;,
    \end{equation}
where $\gamma_{JZ}(t)$ and $\eta(t)$ are some unknown functions of time. However, we are working in a gauge where $A(\tau)$ vanishes so that
    \begin{equation}
        \eval{\pdv{K}{y}}_{y=0} = 2i\eta(t)\tau = 0 \;.
    \end{equation}
Thus $\eta(t)$ is pure gauge and can be set to zero. In this gauge we have
    \begin{equation} \label{gaussansatz}
        K(\tau,\tau') = e^{-\frac{\gamma_{JZ}(t)}{2} (\tau-\tau')^2} \;.
    \end{equation}
A conditional wavefunction that leads to the Gaussian form for $K(\tau,\tau')$ in equation (\ref{gaussansatz}) is of the form,
    \begin{equation} \label{boltz}
        \phi(q,\tau) = \rec{(2\pi\sigma^2)^{1/4}} \exp{-\frac{q^2}{4\sigma^2} + i\frac{q\tau}{\sigma}\sqrt{\frac{\gamma_{JZ}(t)}{\hbar}}} \;,
    \end{equation}
for some real width parameter $\sigma$. If $q$ is understood as a momentum coordinate then this wavefunction could be interpreted as corresponding to a gaseous environment with particle mass $M$ initially at a finite temperature $T$ obeying Boltzmann statistics, in which case,
    \begin{equation}
        \sigma^2 = Mk_BT.
    \end{equation}
Strictly speaking, such an environment would actually be composed of many particles with wavefunctions of the form (\ref{boltz}) and normally distributed average momenta.
The second term in the exponent of (\ref{boltz}) could be interpreted as the effect of a scattering processing - which would be sensitive to the particle position - inducing phase changes in the wavefunctions of scattered environmental particles.

We also assume a Gaussian ansatz for the marginal wavefunction,
    \begin{equation}
        a(\tau, t) = \exp{\frac{1}{2}\delta_{JZ}(t) - (\alpha_{JZ}(t)+i\beta_{JZ}(t))\tau^2} \;,
    \end{equation}
where $\alpha_{JZ}, \beta_{JZ},$ and $\delta_{JZ}$ are functions of time. Since $\delta_{JZ}$ enforces normalization its time dependence will be totally determined by that of $\alpha_{JZ}$. We define the marginal density matrix to be
    \begin{align}
    \begin{split}
        \rho_M(\tau,\tau',t) &= \conj{a}(\tau',t)a(\tau,t) \\ 
                             &= \exp{\delta_{JZ} - \rec{2}\alpha_{JZ}(z^2+y^2) + i \beta_{JZ} yz} \;.
    \end{split}
    \end{align}
Substituting this into the JZME using $\rho_S = \rho_MK$ and equating coefficients of different orders of the coordinates yields the following equations of motion for each parameter,
    \begin{align}
    	& \dot{\delta}_{JZ} = \frac{2\hbar}{m} \beta_{JZ} \\
	    & \dot{\alpha}_{JZ} = \frac{4\hbar}{m} \alpha_{JZ} \beta_{JZ} \label{z2} \\ 
	    & \dot{\gamma}_{JZ} = \frac{4\hbar}{m} \beta_{JZ} \gamma + \frac{2\Lambda}{\hbar} \label{y2} \\
	    & \dot{\beta}_{JZ} = \frac{2\hbar}{m}(\beta_{JZ}^2 - \alpha_{JZ}^2 - \alpha_{JZ} \gamma_{JZ}) \label{yz} \;.
	\end{align}
Defining $1/G(t)=\alpha_{JZ}(t)$ then (\ref{z2}) can be written as
	\begin{equation}
		\beta_{JZ} = -\frac{m}{4\hbar} \frac{\dot{G}}{G} \;,
	\end{equation}
which can be substituted into (\ref{y2}),
	\begin{equation}
		\dot{\gamma}_{JZ}=-\frac{\dot{G}}{G}\gamma_{JZ} + 2\frac{\Lambda}{\hbar} \;,
	\end{equation}
which has a solution for $\gamma_{JZ}$,
	\begin{equation} \label{gammasol}
		\gamma_{JZ}(t) = \frac{2\Lambda}{\hbar}\rec{G(t)}\int_0^t\dd{t'}G(t') \;.
	\end{equation}
With these relationships and the equations of motion (\ref{z2}-\ref{yz}) we find $G(t)$ satisfies
    \begin{equation}
        \dv[3]{t} G(t) = 16\frac{\Lambda\hbar}{m} \;. 
    \end{equation}
Thus $G(t)$ is a cubic polynomial whose coefficients are given by the initial conditions. This agrees with Joos and Zeh's expression for the ensemble width up to a factor of two, which is due to us working with the wavefunction rather than the density matrix. The general solution for $G$ can be written
    \begin{equation}
        G(t) = c_0 + c_1t + c_2t^2 + \frac{8}{3}\frac{\hbar \Lambda}{m}t^3 \;.
    \end{equation}
Using equations (\ref{z2}-\ref{gammasol}), we find the coeffcients can be written,
    \begin{align}
        c_0 &= \frac{1}{\alpha_{JZ}(0)} \\
        c_1 &= -\frac{4\hbar}{m} \frac{\beta_{JZ}(0)}{\alpha_{JZ}(0)} \\
        c_2 &= \frac{4\hbar^2}{m^2}\left(\alpha_{JZ}(0) + \frac{\beta_{JZ}(0)^2}{\alpha_{JZ}(0)}\right) \;.
    \end{align}

\begin{figure*}%
    \label{fig:gammas}
    \centering
    \subfigure{%
        \label{fig:linapproxs}%
        \includegraphics[width=0.47\textwidth]{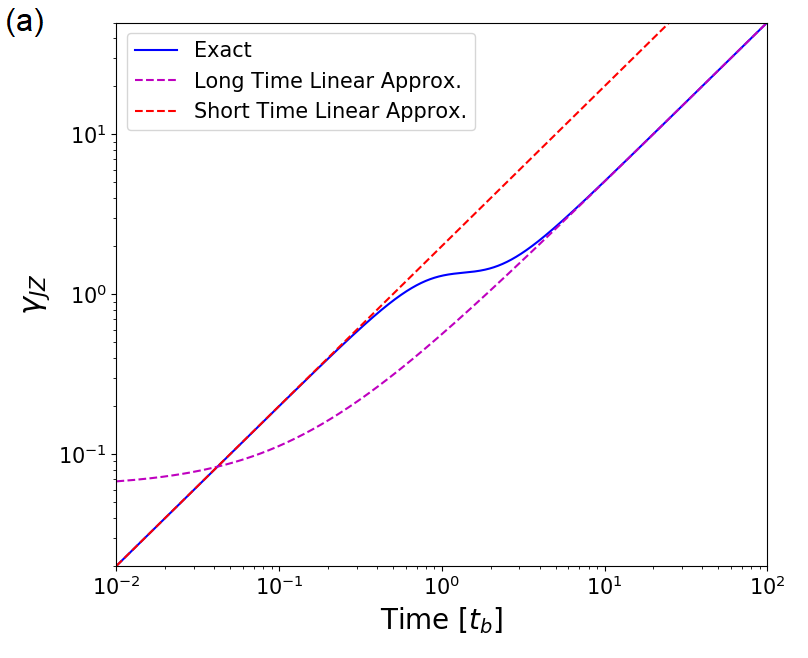}
        }%
    \quad
        \subfigure{%
        \label{fig:coherence_apps}%
        \includegraphics[width=0.45\textwidth]{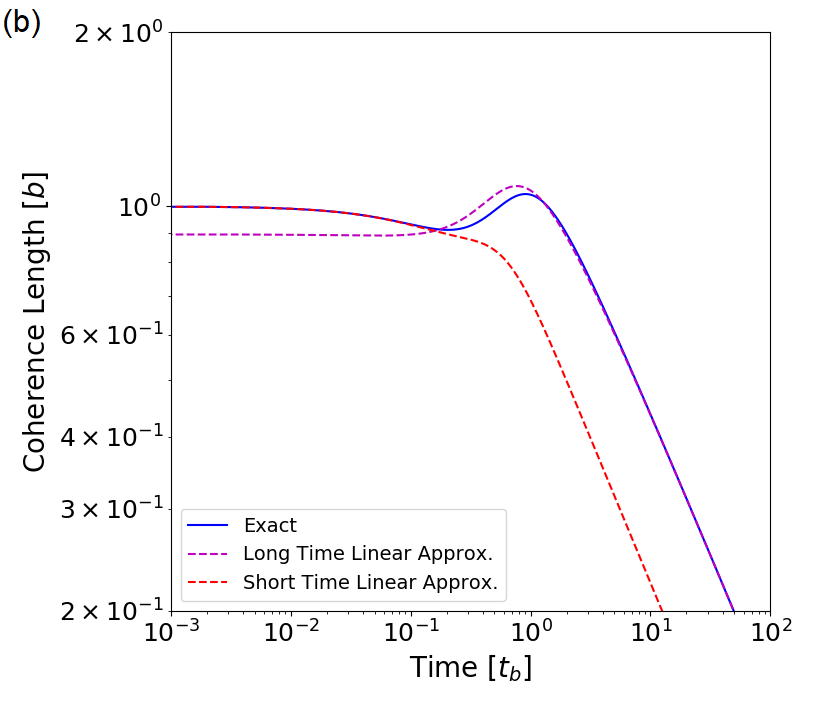}
        }%
    \caption{ (a) The approximations (\ref{gammalin1}) and (\ref{gammalin2}) compared with the exact solution for $\gamma_{JZ}(t)$. (b) The coherence lengths predicted by the linear approximations are compared to the exact solution.}
\end{figure*}

\subsection{Linear Approximation}
Equation (\ref{gammasol}) suggests that we can approximate $\gamma_{JZ}(t)$ by a linear function, since $G$ is a cubic so its integral is quartic. Evaluating the quotient yields
    \begin{equation} \label{gammalin1}
        \gamma_{JZ}(t) = \frac{c_2}{16} + \frac{\Lambda t}{2\hbar} + O(t^{-1}),
    \end{equation}
so $\gamma_{JZ}$ is asymptotically linear in the long-time limit. For short times, $\gamma_{JZ}$ can be Taylor expanded to first order, yielding
    \begin{equation} \label{gammalin2}
        \gamma_{JZ}(t) = \frac{2\Lambda t}{\hbar} + O(t^2).
    \end{equation}
We also show in section (\ref{sec:nonlineareom}) that a $\gamma_{JZ}$ of the second form can be used to yield a non-linear equation of motion for the marginal wavefunction directly. This suggests that it may be possible to effectively capture the influence of the environment on the central system by truncating the equations of motion for the marginal wavefunction at low orders in time. The degree of error this truncation introduces is examined numerically in the following section.

To make this discussion more concrete, we will adopt the initial conditions,
    \begin{align}
    \begin{split}
        &\alpha_{JZ}(0) = \frac{1}{4b^2} \\
        &\beta_{JZ}(0) = 0,
    \end{split}
    \end{align}
corresponding to a real Gaussian wavefunction of width $b$. With these intial conditions, the approximation in (\ref{gammalin2}) becomes accurate to third order in time. The polynomial $G(t)$ thus becomes
    \begin{equation}
        G(t) = 4b^2 + \frac{\hbar^2}{m^2 b^2}t^2 + \frac{8}{3}\frac{\hbar \Lambda}{m}t^3  \;.
    \end{equation}
The coherence length,  $l(t)$ - here defined as the off-diagonal width of the reduced density matrix, $\rho_S = \rho_M K$ - is related to $\gamma_{JZ}(t)$ and $\alpha_{JZ}(t)$ through
    \begin{equation}
        l(t) = \frac{1}{\sqrt{\alpha_{JZ}(t)+\gamma_{JZ}(t)}}.
    \end{equation}
This can be rewritten in terms of the cubic $G(t)$ as 
    \begin{equation} \label{Gint}
        l(t) = \sqrt{\frac{G(t)}{1+\tfrac{2\Lambda}{\hbar}\int \dd{t'} G(t')}} \;.
    \end{equation}
For long times, this approaches
    \begin{equation} \label{coherencelong}
        l(t) = \sqrt{\frac{\hbar}{2\Lambda t}} + O(t^{-1}) \;,
    \end{equation}
whereas for short times
    \begin{equation} \label{coherenceshort}
        l(t) = b-\frac{4\Lambda}{\hbar}b^3t + O(t^2) \;.
    \end{equation}
These agree exactly with the long- and short-time expansions for the coherence length derived by Joos and Zeh \cite{Joos1985} from their treatment of the density matrix. It is interesting to note that for short times, the rate of decay of the coherence length is strongly dependent on the initial width of the Gaussian packet, whereas on long timescales it becomes completely independent of initial conditions. The transition between these two regimes occurs around the characteristic localisation timescale, $t_b$, given by
    \begin{equation}
        t_b = \frac{\hbar}{\Lambda b^2} \;.
    \end{equation}
This can be interpreted as the timescale over which the particle decoheres if we ignored the internal dynamics of the particle (i.e., the first term on the RHS of (\ref{JZME})).

\subsection{Numerical Results}
\label{subsec:numerical_results}

\begin{figure*}%
    \label{fig:linear}
    \centering
    \subfigure{%
        \label{fig:coherence}%
        \includegraphics[width=0.47\textwidth]{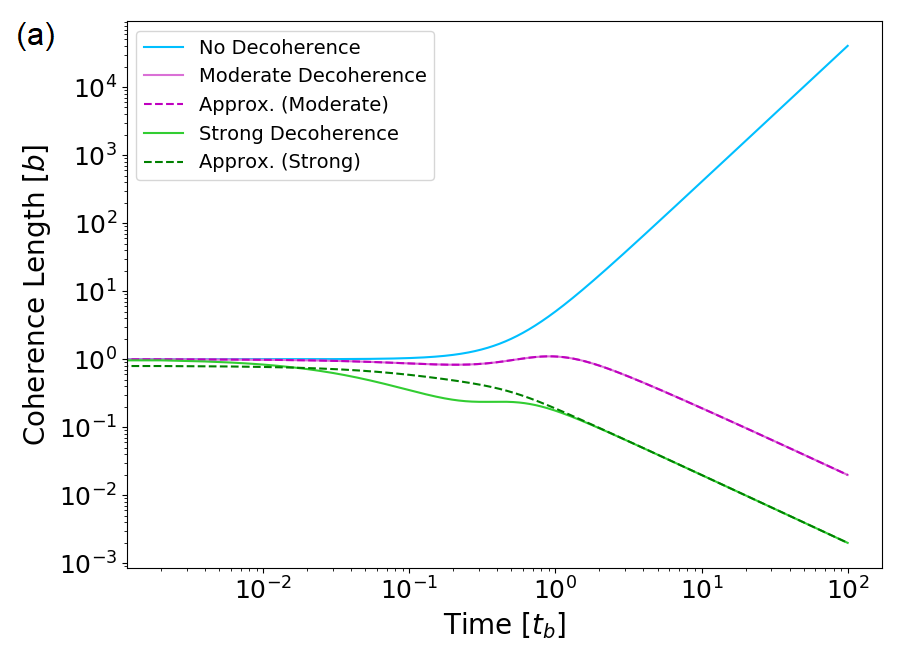}
        }%
    \quad
        \subfigure{%
        \label{fig:ensemble}%
        \includegraphics[width=0.47\textwidth]{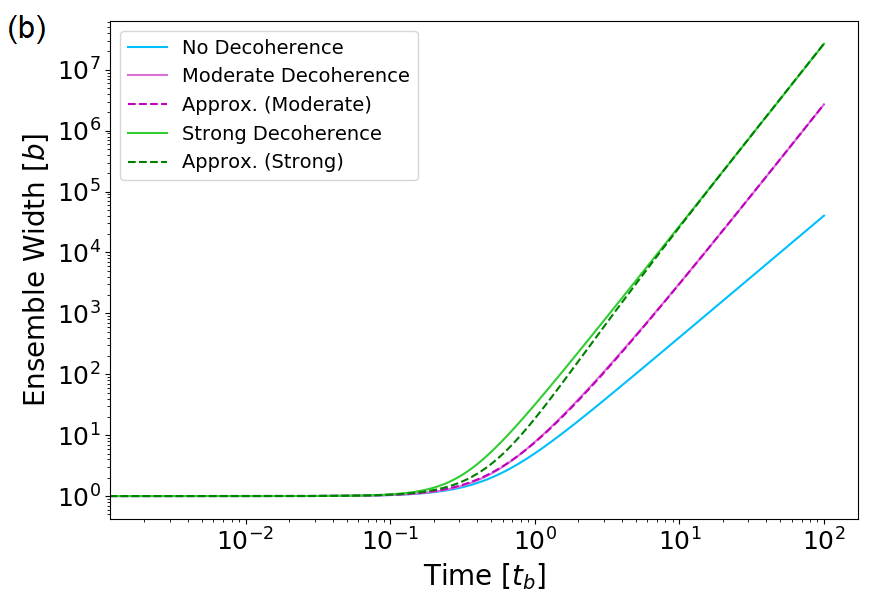}
        }%
    \caption{The time-evolution of (a) the coherence length and (b) the ensemble width for a Gaussian wave packet comparing the exact solution with the long-time linear approximation. For the moderate decoherence $\Lambda = 1/m  $ and for strong decoherence $\Lambda = 10/m $ in units where $b=\hbar=1$.}
\end{figure*}

Figure \ref{fig:coherence} shows the time-evolution of the coherence length for the linear approximation and compares it to the exact solution. Note that the exact solution is produced by both the density matrix formalism and the conditional wave theory, while the linear approximation is motivated by conditional wave theory. We consider a case of moderate decoherence where $m\Lambda b^4/\hbar^2 \sim 1$ and strong decoherence where $m\Lambda b^4/\hbar^2 \sim 10$ . For both moderate and strong decoherence, the linear approximation displays very good agreement with the coherence length of the exact solution in limits at long times and is reasonable agreement at short times. The discrepancy at short times is due to the necessarily strong dependence of the coherence length on the initial condition, as shown by (\ref{coherenceshort}). After approximately one characteristic time interval, the coherence length `forgets' about the initial condition (see (\ref{coherenceshort})), which is consistent with the initial condition independence of the linear term in (\ref{gammalin1}). 

There are also deviations between the linear approximation and the exact solution in the predicted ensemble width near the characteristic coherent length which become stronger with increasing decoherence as shown in  Figure \ref{fig:ensemble}. There is, however, very close agreement in the prediction of the ensemble width at both short and long times.

\section{Non-linear Equation of Motion for the Marginal Wavefunction} \label{sec:nonlineareom}

If we substitute (\ref{gaussansatz}) into the JZME, we find 
    \begin{equation} \label{marginalme}
        \dot{\rho}_m = \frac{2\hbar i}{m} \pdv[2]{\rho_m}{y}{z} - \frac{2\hbar i}{m}\gamma_l(t)y\pdv{\rho_m}{z} + \left(\frac{\dot{\gamma_l}}{2}- \frac{\Lambda}{\hbar} \right) y^2 \rho_m \;,
    \end{equation}
where we have renamed $\gamma(t)=\gamma_l(t)$ to distinguish it from the function in the previous section.
This can be reproduced from the equation of motion for the marginal wavefunction (\ref{aeom}) provided that
    \begin{align}
        \gamma_l(t) &= \frac{2\Lambda}{\hbar}(t-t_0) \label{gauge1} \\ 
        \eps(\tau,t) &= \frac{2\hbar\Lambda}{m} (t-t_0) \ln \abs{a(\tau)}^2 \;, \label{lseeps}
    \end{align}
so that the equation of motion for $a$ reads
    \begin{equation} \label{lse}
        i\dot{a} = -\frac{\hbar}{2m} \laplacian a + \frac{2\hbar\Lambda}{m} (t-t_0) a \ln \abs{a(\tau)}^2 \;.
    \end{equation}
That equations (\ref{lse}) and (\ref{marginalme}) are equivalent is shown in Appendix \ref{sec:appendix2}. Equation \eqref{lse} takes the form of the logarithmic Schr{\"o}dinger equation where the magnitude of the non-linear term increases linearly with time. The logarithmic Schr{\"o}dinger equation has bounded solutions when the sign of the interaction is negative \cite{Bialynicki1979}. In the decoherence case the sign is positive and does not produce bounded solutions, consistent with the results discussed in Section \ref{subsec:numerical_results}.

\section{Conclusion}

Conditional wave theory provides a unique and interesting approach to decoherence, because it involves calculations on wave functions rather than density matrices. The factorization at the heart of CWT enables the density matrix to be separated into a term that describes the particle's spatial density, and the gauge terms that contain information about the entanglement with the environment. We have shown that in the general case the number of environmental terms required is infinite and contain equivalent information to the particle's reduced density matrix. Nevertheless, the CWT approach enables approximations that specifically target the environmental terms, which could be exploited for computational advantages or to gain new insights into how decoherence affects the time-evolution of the particle's density. Motivated by analytic Gaussian solutions of the JZ model of collisional decoherence, we introduce a linear time approximation and showed that it gives very close agreement to the exact solution in the limits of short and long times, but a discrepancy near the characteristic decoherence time. We note, however, that the JZ model ignores the effects of recoil and dissipation, and does nessarily provide an accurate model of physical systems at long times. This linear approximation can potentially open up opportunities for computationally efficient simulations of non-Gaussian wave functions. It is highly interesting that the linear-approximation leads to a logarithmic Schr{\"o}dinger equation for the particle's marginal wave function, which may also lead to new physical insights into the dynamics of environmental interactions. The CWT formalism is general and we envision extensions to model decoherence in systems with more than one particle and with discrete degrees of freedom. 

\appendix

\section{g-Functions}
\label{g-funcs}
In the case of continuous system coordinates, integrals of the form
    \begin{equation}
        g_{(n,m)}(\vtau) = \int \dd{\vb{r}} (\gratau^n\conj{\phi})\vdot(\gratau^m\phi) 
    \end{equation}
frequently appear in CWT, and some of their properties and relations are discussed here.
For even $N=n+m$, $g_{(n,m)}(\vtau)$ is a complex scalar function whereas for odd $N$ it is a complex vector field. If we adopt the convention that a $0^{\mathrm{th}}$ derivative is the identity then (\ref{phinorm}) can be written as,
    \begin{equation} \label{g00}
        g_{(0,0)}(\vtau) = 1,
    \end{equation}
for all $\vtau$, that is, it is a constant function.
Furthermore, it follows from the product rule that (where the differentiation is understood as a gradient or divergence as appropriate)
    \begin{equation}
        \gratau g_{(n,m)}(\vtau) = g_{(n+1, m)}(\vtau) + g_{(n, m+1)}(\vtau).
    \end{equation}
Therefore, differentiating (\ref{g00}) $N$ times we obtain
    \begin{equation} \label{binom}
        \sum_{k=0}^N \binom{N}{k} g_{(N-k,k)}(\vtau) = 0.
    \end{equation}
From the definition of these g-functions we also observe that
    \begin{equation}
        g_{(m,n)}(\vtau)=g_{(n,m)}^{*}(\vtau) \;,
    \end{equation}
so if we expand (\ref{binom}) for $N=1$ we find
    \begin{equation} \label{g01}
        g_{(0,1)}(\vtau)=-g_{(1,0)}(\vtau) = - g_{(0,1)}^{*}(\vtau) \;,
    \end{equation}
which implies $-ig_{(0,1)}(\vtau)$ is purely real. In fact it is clear by inspection that $\vb{A}(\vtau)=-i\hbar g_{(0,1)}(\vtau)$ and so the vector part of the gauge field is real. Expanding (\ref{binom}) to second order, we find
    \begin{equation}
        g_{(0,2)}(\vtau) + 2g_{(1,1)}(\vtau) + g_{(2,0)}(\vtau) = 0
    \end{equation}
and noting that $g_{(n,n)}(\vtau)$ is real for all $n$, we find $\Re{g_{(0,2)}(\vtau)} =-g_{(1,1)}(\vtau)$. Similar constraints can be found at all orders of the g-functions.
We now wish to calculate how many of the g-functions are independent and how these terms are related to the Taylor series expansion of the density matrix. Consider the 1D case. Let the set $g_N = \{ g_{(n,m)}(\vtau) | N = n + m \}$. The elements of $g_N$ can be written in terms of $N+1$ real scalar functions. One of these scalar functions is not independent of the others because of Eq. \eqref{binom}. Of the remaining $N$ functions, $N-1$ can be written as derivatives of terms in $g_{N-1}$. It is then possible to show by induction that the set $\{ g_1, ..., g_N \}$ can be written in terms of $N$ independent real scalar functions. This result proves to be important in understanding the equivalence of the CWT and density matrix approaches to decoherence.

\section{Time Evolution of the Gauge Field}
\label{sec:appendix2}
    Using equation (\ref{phieom}) we can immediately write down the time evolution of $\vb{A}$ as
        \begin{align}
        \begin{split} \label{Adot}
            &\dot{\vb{A}}(\tau) = -i\hbar \int \dd{r} \left(\dot{\conj{\phi}}\gratau\phi +     \conj{\phi}\gratau\dot{\phi}\right) \\
            &= \int \dd{r} \big( ( H_{\phi}^{\dagger}\conj{\phi} - \eps\conj{\phi}) \gratau\phi - \conj{\phi}\gratau(H_{\phi}\phi - \eps\phi) \big)
        \end{split}
        \end{align}
        and define the conditional expectation value of the energy operator as
        \begin{equation}
            A_0 = i\hbar \cev{\pdv{t}} \;.
        \end{equation}
    The term $\cev{H_\phi}$ is independent of the marginal wavefunction (since it couples through the $D_-$ term, which has a vanishing conditional expectation value). Thus (\ref{Adot}) can be written,
        \begin{equation}
            \dot{\vb{A}}-\gratau A_0 = 2 \Re \braket{\gratau \phi}{H_\phi \phi}_E \;.
        \end{equation}
    The right hand side will have a term proportional to $g_{(1,2)}(\vtau)$, due to the presence of the Laplacian in $H_\phi$. Thus to find the time evolution of $A\propto g_{(0,1)}(\vtau)$ we need to find the time evolution of $g_{(1,2)}(\vtau)$, which is going to depend on yet higher order $g$-functions by precisely the same argument. 
    
\section{Equivalence between JZME solution subset and solutions to the modified LSE}
\label{sec:appendix3}

    Beginning with the 1-dimensional marginal equation of motion, (\ref{aeom}) in a gauge where $A=0$,
        \begin{equation}
            i\dot{a}(\tau,t) = -\frac{\hbar}{2m} \pdv[2]{a(\tau,t)}{\tau} + \rec{\hbar}\eps(\tau)a(\tau,t) \;,
        \end{equation}
    we can construct the corresponding equation of motion for the marginal density matrix,
        \begin{equation} 
            i\dot{\rho}_m = -\frac{\hbar}{2m}\left(\partial^2_\tau - \partial^2_{\tau'}\right) \rho_m + \frac{1}{\hbar}\left[ \eps(\tau)-\eps(\tau') \right] \rho_m \;.
        \end{equation}
    To match this with equations (\ref{marginalme}) and (\ref{gauge1}), we require
        \begin{equation} \label{req}
            \frac{4\Lambda}{m}(t-t_0)(\tau-\tau')\pdv{\rho_m}{z} = \frac{1}{\hbar}\left[ \eps(\tau)-\eps(\tau') \right] \rho_m.
        \end{equation}
    Noting that
        \begin{equation}
            \pdv{\rho_m}{z} = \frac{1}{2}\left(\conj{a}(\tau') \pdv{a(\tau)}{\tau} + a(\tau) \pdv{\conj{a}(\tau')}{\tau'}\right)  \;,
        \end{equation}
    we can rewrite (\ref{req}) as
        \begin{equation}
            \frac{\eps(\tau)-\eps(\tau')}{\tau-\tau'} = \frac{2 \Lambda t'}{m}\left( \frac{1}{\conj{a}(\tau')} \pdv{\conj{a}(\tau')}{\tau'} + \frac{1}{a(\tau)} \pdv{a(\tau)}{\tau} \right).
        \end{equation}
    where $t'=t-t_0$. Taking the limit as $\tau' \rightarrow \tau$ yields
        \begin{align}
            \pdv{\eps(\tau)}{\tau} = \pdv{\tau}\left( \frac{2\Lambda t'}{m} \ln |a|^2 \right) \;,
        \end{align}
    which can be integrated with respect to $\tau$  to yield (\ref{lseeps}). 
    
\bibliography{main.bib}
\bibliographystyle{unsrt}

\end{document}